\documentclass[twocolumn]{aastex631}

\def\arcsec{{^{\prime\prime}}}

\def\farcs{\hbox{$.\!\!^{\prime\prime}$}}

\def\gtrsim{\mathrel{\hbox{\rlap{\hbox{\lower4pt\hbox{$\sim$}}}\hbox{$>$}}}}
\def\lessim{\mathrel{\hbox{\rlap{\hbox{\lower4pt\hbox{$\sim$}}}\hbox{$<$}}}}

\usepackage{xspace}
\newcommand{\galario}{\texttt{galario}\xspace}
\newcommand{\frank}{\texttt{frank}\xspace}
\newcommand{\EDDY}{\texttt{EDDY}\xspace}
\newcommand{\tclean}{\texttt{TCLEAN}\xspace}
\newcommand{\bettermoments}{\texttt{bettermoments}\xspace}
\newcommand{\uv}{\texttt{uv}\xspace}

\begin{document}
\title{Hints of Disk Substructure in the First Brown Dwarf with a Dynamical Mass Constraint}

\received{March 7, 2025}
\accepted{May 11, 2025}
\correspondingauthor{Alejandro Santamar\'ia-Miranda}
\email{alejandrosantamariamiranda@gmail.com}

\author[0000-0001-6267-2820]{Alejandro Santamar\'ia-Miranda}
\affiliation{Departamento de Astronom\'ia, Universidad de Chile, Camino El Observatorio 1515, Las Condes, Santiago, Chile }
\author[0000-0003-2045-2154]{Pietro Curone}
\affiliation{Departamento de Astronom\'ia, Universidad de Chile, Camino El Observatorio 1515, Las Condes, Santiago, Chile }
\author{Laura P\'erez}
\affiliation{Departamento de Astronom\'ia, Universidad de Chile, Camino El Observatorio 1515, Las Condes, Santiago, Chile }
\author{Nicolas T. Kurtovic}
\affiliation{Max-Planck-Institut für Extraterrestrische Physik, Giessenbachstrasse1, 85748 Garching, Germany}
\author[0000-0002-7238-2306]{Carolina Agurto-Gangas}
\affiliation{Departamento de Astronom\'ia, Universidad de Chile, Camino El Observatorio 1515, Las Condes, Santiago, Chile }
\author{Anibal Sierra}
\affiliation{Mullard Space Science Laboratory, University College London, Holmbury St Mary, Dorking, Surrey RH5 6NT, UK}
\author{Itziar de Gregorio-Monsalvo}
\affiliation{European Southern Observatory, 3107, Alonso de C\'ordova, Santiago de Chile}
\author{Nuria Hu\'elamo}
\affiliation{Centro de Astrobiolog\'ia (INTA-CSIC); ESAC campus, Camino bajo del Castillo s/n, Urb. Villafranca del Castillo, 28692 Villanueva de la Ca\~nada, Madrid, Spain}
\author{James M. Miley}
\affiliation{Departamento de F \'isica, Universidad de Santiago de Chile, Av. Victor Jara 3659, Santiago, Chile}
\affiliation{Millennium Nucleus on Young Exoplanets and their Moons (YEMS)}
\affiliation{Center for Interdisciplinary Research in Astrophysics Space Exploration (CIRAS), Universidad de Santiago de Chile, Chile}
\author{A\'ina Palau}
\affiliation{Instituto de Radioastronom\'ia y Astrof\'isica, Universidad Nacional Aut\'onoma de M\'exico, Antigua Carretera a P\'atzcuaro $\#$8701, Ex-Hda. San Jos\'e de la Huerta, Morelia, Michoac\'an, C.P. 58089, M\'exico}
\author{Paola Pinilla}
\affiliation{Mullard Space Science Laboratory, University College London, Holmbury St Mary, Dorking, Surrey RH5 6NT, UK}
\author{Isabel Rebollido}
\affiliation{European Space Agency (ESA), European Space Astronomy Centre (ESAC), Camino Bajo del Castillo s/n, 28692 Villanueva de la Cañada, Madrid, Spain}
\author{\'Alvaro Ribas}
\affiliation{Institute of Astronomy, University of Cambridge, Madingley Road, Cambridge, CB3 0HA, UK}
\author{Pablo Rivi\`ere-Marichalar}
\affiliation{Observatorio Astron\'omico Nacional (OAN,IGN), Calle Alfonso XII, 3. 28014 Madrid, Spain}
\author{Matthias R. Schreiber}
\affiliation{Departamento de F\'isica, Universidad T\'ecnica Federico Santa Mar\'ia, Av. España 1680, Valpara\'iso, Chile}
\author[0000-0003-4361-5577]{Jinshi Sai}
\affiliation{Academia Sinica Institute of Astronomy \& Astrophysics, 11F of Astronomy-Mathematics Building, AS/NTU, No.1, Sec. 4, Roosevelt Rd, Taipei 10617, Taiwan, R.O.C.}
\author{Benjam\'in Carrera}
\affiliation{Departamento de Astronom\'ia, Universidad de Chile, Camino El Observatorio 1515, Las Condes, Santiago, Chile }

\begin{abstract}
We present high-resolution ALMA observations at 0.89\,mm of the Class II brown dwarf 2MASS J04442713+2512164 (2M0444), achieving a spatial resolution of 0\farcs046 ($\sim$6.4 au at the distance to the source). These observations targeted continuum emission together with $^{12}$CO\,(3-2) molecular line. The line emission traces a Keplerian disk, allowing us to derive a dynamical mass between  0.043\,-\,0.092\,M${_{\odot}}$ for the central object. We constrain the gas-to-dust disk size ratio to be $\sim$7, consistent with efficient radial drift. However, the observed dust emission suggest that a dust trap is present, enough to retain some dust particles. We perform visibility fitting of the continuum emission, and under the assumption of annular substructure, our best fit shows a gap and a ring at 98.1$^{+4.2}_{-8.4}$\,mas ($\sim$14\,au) and 116.0$^{+4.2}_{-4.8}$\,mas ($\sim$16\,au), respectively, with a gap width of 20\,mas ($\sim$3\,au). To ensure robustness, the data were analyzed through a variety of methods in both the image and \uv plane, employing multiple codes and approaches. This tentative disk structure could be linked to a possible planetary companion in the process of formation. These results provide the first dynamical mass of the lowest mass object to date, together with the possible direct detection of a substructure, offering new insights into disk dynamics and planet formation in the very low-mass regime. Future higher spatial resolution ALMA observations will be essential to confirm these findings and further investigate the link between substructures and planet formation in brown dwarf disks.

\end{abstract}


\section{Introduction} \label{sec:intro}

Brown dwarfs (BDs) are substellar objects that cannot fuse hydrogen in their cores. Interest in BDs has surged over the last decades following the discovery of exoplanets orbiting them, with masses ranging from giant planets \citep{Chauvin2004} to super-Earths \citep{Bennett2008}. BDs are known to harbor circumstellar disks where these planets form with scaled-down properties, more compact, fainter and less massive, compared to disks around T-Tauri stars \citep{Testietal16-1},  making them excellent laboratories for testing planet formation under extreme conditions of low density and temperature.

The discovery of giant planets around BDs challenges existing planet formation theories, particularly those based on planetesimal accretion \citep{Liu2020}. Millimeter flux measurements suggest that even the most massive BD disks contain only a few Jupiter masses of material. According to core accretion models, disks with a total mass of just a few M$_{\mathrm{Jup}}$ surrounding a 0.05 M$_{\odot}$ BD should not be capable of forming giant planets, though they could potentially form rocky planets \citep{Payne07-1}. This holds unless giant planets form at very early stages, as suggested by the abundance of substructures found in young disks around stars \citep{Andrews2020}. Additionally, the presence of millimeter to centimeter-sized particles in disks around very low-mass stars and BDs, which has been inferred from low spectral indices measured by The Atacama Large Millimeter/submillimeter Array (ALMA) \citep{Ricci12-1}, is difficult to reconcile with dust evolution models due to the radial drift barrier. Radial drift causes dust particles to migrate inward towards the central star as they grow within the disk, driven by the sub-Keplerian velocity of the gas \citep{Weiden77}. In BD disks, radial inward drift velocities are expected to be higher than in stellar disks, leading to the depletion of large grains that cannot continue to grow \citep{Pinilla2017}. Consequently, BD disks would not be expected to contain millimeter-sized grains, and planet formation should be inhibited—contradicting both exoplanet discoveries around BDs and ALMA observations of BD disks. To account for these submillimeter observations, radial drift in the disks around brown dwarfs must be mitigated. One plausible explanation is the presence of strong pressure bumps distributed radially within the disk \citep{Pinilla2013, Pinilla2017}. These bumps could create substructures, such as rings, similar to those observed in higher-mass objects (e.g. \citealt{Andrews2020}).

These types of substructures have been ubiquitously observed in stars of all masses. Substructures have also been detected in young massive stars \citep[8-38 M$_{\odot}$]{Frost21} as well as in very low-mass stars (0.1-0.2 M$_{\odot}$, like e.g. large cavities \citep{Pinilla21}, gaps, and rings \citep{Kurtovic21}, or asymmetric dust rings \citep{Hashimoto21}. However, the substellar regime has not been explored, and no substructure has been detected yet. 

One of the brightest brown dwarf disks is 2MASS J04442713+2512164 (hereafter 2M0444), by far the best candidate to study planet formation and circumstellar disk structures in the BD regime: it is an M7.25 spectral type and a Class II source \citep{Luhman10} located in Taurus at a distance of $\sim$140 pc, with L$_{*}$ =0.028 L$_{\odot}$ and M$_{*}$= 0.05 M$_{\odot}$, well into the substellar regime and likely formed as a scaled-down version of low mass stars \citep{Riccietal14-1, ricci17}. This BD disk was observed at different wavelengths, and it is one of the largest disks resolved in the submillimeter by ALMA, with an estimated outer dust disk radius of  $\sim$100 au \citep{Rilinger2019}. It shows evidence of grain growth according to the derived submillimeter spectral slope ($\alpha=2.37\pm0.32$) between ALMA Band 7 (0.9 mm) and Band 3 (3 mm) \citep{ricci17} and advanced dust processing based on the high abundance of crystalline silicates \citep{Bouy2008}. The estimated mass of the disk is near 2 M$_{\mathrm{Jup}}$, capable of forming rocky planets as predicted by core accretion models \citep{Payne07-1}. Additionally,  its spectral energy distribution is better reproduced by a disk with a dust gap \citep{Rilinger2019}, suggesting the presence of substructure(s) within the disk.  

In this Letter we present very high resolution ALMA Band 7 observations (0\farcs046) in the continuum (at 0.89 mm) and ($\sim$0\farcs062) in CO(3-2) line emission of 2M0444.
The Letter is structured as follows. In Section \ref{sec:p10_obs}, we describe the observations and data reduction. In Section \ref{sec:p10_anal}, we present the analysis of the continuum and the gas fitting in both the image and visibility planes. In Section \ref{sec:p10_discussion}, we discuss the hint of substructure, the efficiency of radial drift, and the potential planet mass that may be responsible for the possible substructure. Finally, in Section \ref{sec:p10_summary}, we summarize the main results.

\section{Observation and data reduction}
\label{sec:p10_obs}
2M0444 was observed with the ALMA 12-meter array in Band 7 as part of project 2023.1.00158.S (PI: A. Santamaría-Miranda). Observations using the long baseline configuration (C43-8) were carried out on three dates between October and November 2023, followed by an additional observation in May 2024 using the shorter baseline configuration C43-5. C43-8 was the most extended configuration available in Cycle 10, while C43-5 complemented it by recovering any extended emission filtered out in C43-8. Detailed observational parameters are summarized in Appendix Table \ref{Tab:p10_observations}.

The correlator was set to dual-polarization mode with four basebands. Three basebands were centered at frequencies of 331.362, 333.320, and 343.362 GHz, respectively, with spectral resolutions ranging from 1.36 to 1.41 km s$^{-1}$. The fourth baseband was tuned to the $^{12}$CO (3-2) line at 345.796 GHz, with a spectral resolution of 0.42 km s$^{-1}$. $^{13}$CO (3-2) line at 330.58 GHz was detected at the first baseband. Each baseband had a bandwidth of 1.875 GHz. The total on-source integration time was 2.01 hours for the long-baseline observations and 0.71 hours for the short-baseline configuration.

The data calibration followed the standard ALMA pipeline procedures. Subsequently, we applied self-calibration using version 6.5.4.9 of the Common Astronomy Software Application (CASA) package \citep{McMullin2007}. Prior to self-calibration, each execution was imaged individually, and we aligned the emission peaks using the tasks \textsc{fixvis} and \textsc{fixplanets}. Self-calibration began after flagging the lines and involved five phase-only iterations on the short-baseline data, with the interval time reduced progressively after each iteration. Then, we combined both long- and short-baseline data for joint self-calibration, performing eight phase-only iterations followed by one phase-amplitude iteration. The final self-calibration solutions were applied to both the continuum and spectral line data, and \tclean was used to generate continuum and line images, with continuum subtraction applied beforehand. By doing so, we improved the signal-to-noise (s/n) of the emission peak by 271$\%$ (from 42 to 114). A Briggs robust parameter of 0.5 was selected for the continuum image to balance sensitivity and resolution, while a robust parameter of 2 was used for the $^{12}$CO and $^{13}$CO  emission lines to maximize sensitivity. Before extracting physical properties from the images, a primary beam correction was applied. The angular resolution of the continuum maps is 0$\farcs$046 (equivalent to $\sim$6.4 au at a distance of 140 pc) and $\sim$0\farcs062 ($\sim$8.7 au) for the $^{12}$CO emission line. The accuracy of the flux calibration in Band 7 is estimated to be 10$\%$ \citep{Francis2020}. The continuum and $^{12}$CO emission line maps are shown in Figure \ref{momentos}. 

\section{Analysis}
\subsection{Disk rotation in the image plane}
\label{sec:p10_SLAM}

\begin{figure*}
\includegraphics[width=0.99\textwidth]{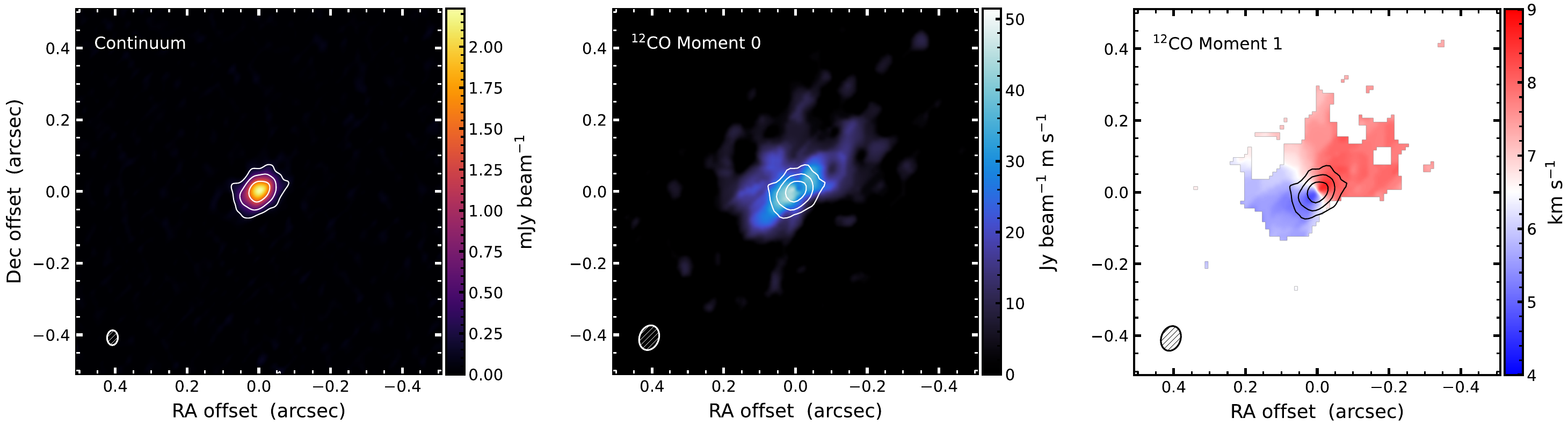}
\caption{Continuum and moment 0 and 1 maps of the $^{12}$CO using a Keplerian mask. The left panel shows the continuum image of 2M0444 at $\sim$0.89 mm reconstructed using Briggs weighting and robust= 0.5.  Center panel shows the $^{12}$CO integrated intensity map, and the right panel shows the $^{12}$CO intensity-weighted velocity map. 
Contours represent the disk continuum emission with 3, 5 and 10 times the rms (1$\sigma$ = 25.8 $\mu$Jy beam$^{-1}$). The beam size is represented by the filled ellipse in the bottom left corner of each panel.}  \label{momentos}
\end{figure*}

The $^{12}$CO(3-2) emission line traces a rotating disk-like structure around 2M0444 (Figure \ref{momentos}). This line is detected from velocities of 3.08 km s$^{-1}$ to 10.28 km s$^{-1}$ with respect to the local standard of rest (LSR), with a s/n exceeding 3$\sigma$ (rms = 2 mJy/beam). The LSR velocity of 2M0444, derived from the spectra over the channel map, is 6.7 km s$^{-1}$ consistent with the Taurus LSR \citep{Dobashi}. The redshifted and blueshifted emission appears to be consistent with Keplerian rotation. To derive the dynamical mass of the central object, we employed \EDDY (Extracting Disk Dynamics) \citep{eddy}.

\EDDY is a Python toolkit designed to efficiently extract precise velocity profiles from Doppler-shifted line emission in protoplanetary disks. It simplifies the process of fitting first moment maps and inferring rotational velocities by analyzing spectra in annular regions of the disk.  We used the \bettermoments package \citep{bettermoments} to generate zeroth and first-order moment maps, which were then input into \EDDY for an initial fit. This provided estimates for key parameters such as the initial position (x$_{0}$, y$_{0}$), position angle (PA), mass of the central object, and the LSR velocity (VLSR). The inclination was fixed using the value derived in Section \ref{sec:p10_cont_fit} which is 49.9 deg, and we assumed a distance of $\sim$ 140 pc \citep{Galli18-2}.

We performed multiple iterations of \EDDY until the parameters converged. Using these results, we created a Keplerian mask \footnote{\url{https://github.com/richteague/keplerian_mask}} \citep{Teague2020}. This mask was applied to the moment maps via \bettermoments to isolate the Keplerian emission. The final refinement of the central mass was then achieved using \EDDY once more, applying the Keplerian mask to the moment 1 map. This process yielded a central mass for 2M0444 of 0.043$^{+0.003}_{-0.002}$M$_{\odot}$, with a VLSR of 6700 $\pm$ 200 m s$^{-1}$, and a position angle of 307.2$\mathrm{^{o}}$ $\pm$ 1.8$\mathrm{^{o}}$. Figure \ref{cornerplot_eddy} in Appendix shows the corner plot of the final parameter distributions obtained from the \EDDY analysis.
The $^{13}$CO(3-2) emission line (Figure \ref{momentos_no_keplerianos} in Appendix) also traces rotation in a more compact disk, but the s/n is too low to derive a reliable estimate of its dynamical mass.

\subsection{Fitting of the visibilities for the $^{12}$CO line emission} \label{sect:fit_12co}
We fit the visibilities of the $^{12}$CO emission following a similar procedure to that of \citet{kurtovic2024} in MHO\,6. 
The visibilities were extracted channel by channel, and we fitted 30 channels centered at the source systemic velocity. The brightness model cube considered two emitting layers to recreate the $^{12}$CO emission: A front and back emitting layer, each one with its own independent height and temperature profile. The model was fitted using MCMC (Markov Chain Monte Carlo), with 192 walkers (equal to eight times the number of free parameters), and a flat prior for all the parameters. The only additional condition imposed on the model was that the back layer could not be at a larger distance from the midplane than the front layer. 
The MCMC  achieved convergence after about $10^4$ steps, and we continued running for $10^4$ steps to sample the parameter space. 

The brightness distribution of the highest likelihood model for the $^{12}$CO emission is shown in Figure \ref{app:fig:vismodel_12co} in Appendix. The visibilities of this model were subtracted from those of the observation, and the image of those residuals is also shown in the same gallery. After subtracting the model. The uncertainty for each parameter is measured from the dispersion of the walkers. For the geometrical parameters, we find consistent solutions of inclination and position angle to those from the continuum. For the stellar mass, we obtain $M_\star=0.092_{-0.007}^{+0.019}$ M$_{\odot}$, where the uncertainty represents the $99.7$ percentile distribution, the equivalent to $3\sigma$. The same brightness model can be used to measure other disk properties, such as the disk size, as discussed in Section \ref{sec:gas-dust-size}. 

\label{sec:p10_anal}
\subsection{Fitting of the visibilities for the continuum}

\label{sec:p10_cont_fit}

\begin{figure*}
\includegraphics[width=\textwidth]{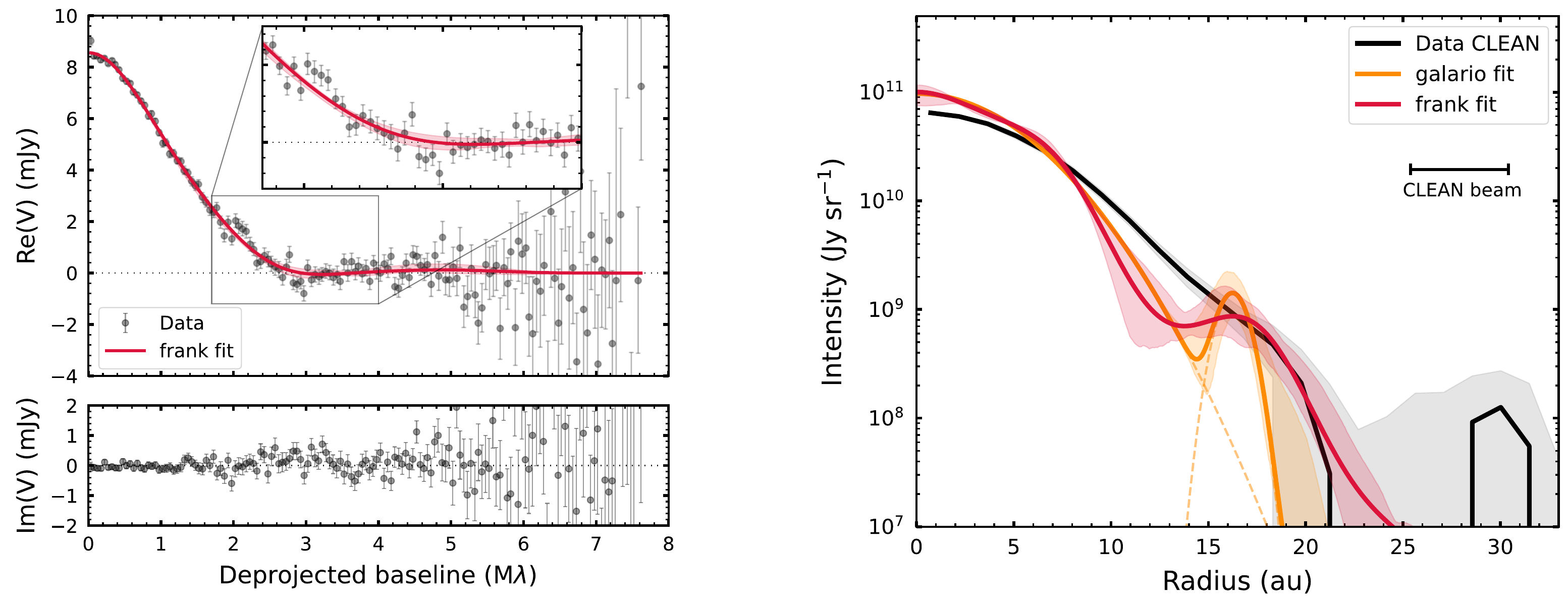}
\caption{Left: Real and imaginary parts of the recentered and deprojected visibilities, azimuthally averaged into 50 k$\lambda$-wide bins, shown as a function of the deprojected baseline length for the data (gray points) and the best-fit models from \frank (red line). The inset box highlights the region where substructures in the observed data are more prominent. Note that the \frank best-fit model is shown only for the real part, as the imaginary part is not fitted due to the assumption of an axisymmetric model. Right: Intensity radial profiles of the azimuthally averaged CLEAN data with robust 0.0 (black), along with the best-fit models from \galario (orange) and \frank (red). The dashed orange lines indicate the best fit of the Gaussian components assumed in the \galario model. The red and orange shadings indicate the 16th and 84th percentiles from the \frank bootstrapping and the \galario MCMC marginalized distribution, respectively. The black scale bar in the right panel shows the average between the major and minor axes of the FWHM CLEAN synthesized beam.}
\label{fig:uv_lambda}
\end{figure*}

\begin{figure*}
\includegraphics[width=\textwidth]{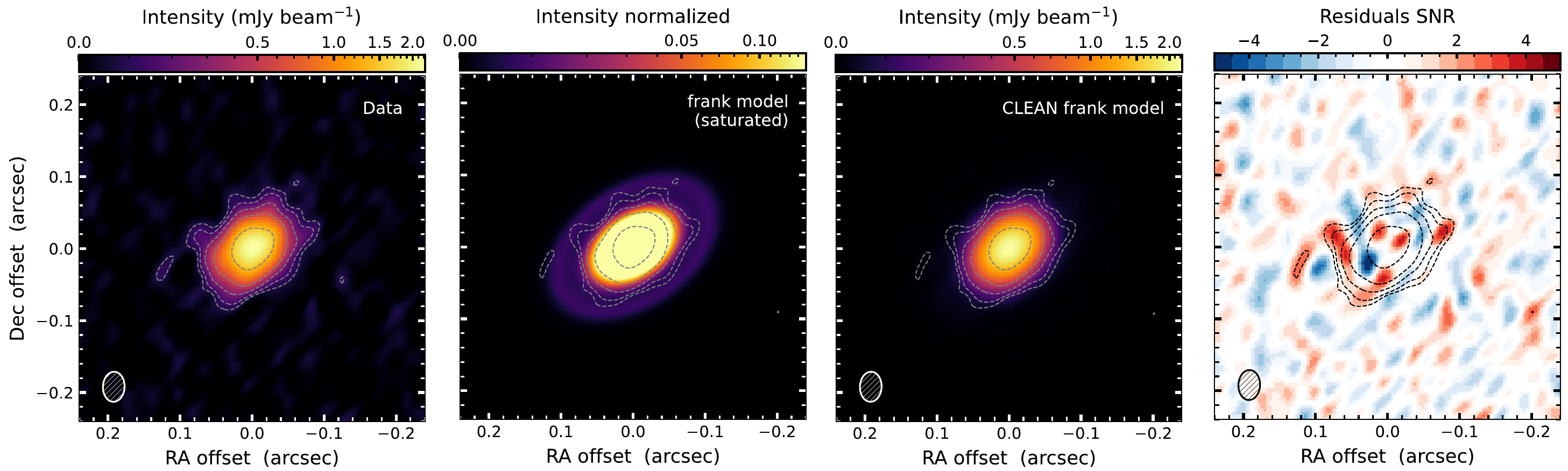}
\caption{CLEAN data, \frank model, \frank CLEAN image, and residual between data and \frank model imaged with robust 0.0. Dashed contours indicate the [3, 5, 10, 20, 50]$\sigma$ levels of the CLEAN data image.}
\label{fig:continuum_gallery}
\end{figure*}

Before fitting the visibilities associated with the dust continuum emission of 2M0444, we subtracted the visibilities associated with an external source within the field of view using the CASA task \texttt{uvsub}. This external source, located ${\sim}2.6\arcsec$ ($\sim$364 au) to the northeast of 2M0444, has an integrated flux density of ${\sim}1.3$~mJy. For the imaging, we employed the \texttt{tclean} task in CASA with  multiscale deconvolver, using scales of [0, 8, 15, 30] pixels, iterating down to a $1\sigma$ threshold with a central circular mask of $0.3\arcsec$ in radius.

To characterize the morphology of the continuum emission of 2M0444, we utilized a two-step pipeline for visibility fitting. First, we applied the code \galario \citep{Tazzari2018} to perform a parametric fit which was solely aimed at retrieving the disk’s geometrical parameters, including inclination ($i$), position angle (PA), and offsets in right ascension ($\Delta$RA) and declination ($\Delta$Dec) between the disk center and the observation phase center. We determined the best-fit model by minimizing the $\chi^2$ value, sampling the parameter space with an MCMC approach using the \texttt{emcee} package \citep{emcee}. For the disk’s parametric model, we assumed an axisymmetric emission profile composed of a central Gaussian and an additional Gaussian ring with the function $I(R) = f_0 \exp(-R^2 / 2\sigma_0^2) + f_1 \exp(-(R - R_1)^2 / 2\sigma_1^2)$. Uniform priors were employed, with the intensity normalization factors $f0$ and $f1$ sampled logarithmically. The derived geometrical parameters are summarized in Table \ref{Tab:geom_par}.

Following this, we used the obtained geometrical parameters as inputs for a second visibility fitting with the code \texttt{frankenstein} (\citealt{Jennings2020}, hereafter \frank). The frank code reconstructs the azimuthally averaged brightness profile of a source as a function of disk radius by directly fitting the real component of the deprojected, unbinned visibilities against the baseline. The brightness profile is derived non-parametrically by fitting the visibilities with a Fourier-Bessel series, with Gaussian process regularization. The primary advantage of \frank is its ability to retrieve sub-beam resolution features in the dust continuum emission while leveraging the full data sensitivity, all without assuming a parametric model. Both \galario and \frank have been successfully applied to characterize the emission of compact disks around higher mass objects (e.g., \citealt{Facchini19}, \citealt{Jennings2022}, \citealt{Curone2023}), but this study marks their first application to a previously reported brown dwarf disk. After testing several values of the \frank hyperparameters, we selected conservative values of $\alpha=1.20$ and $w_\mathrm{smooth}=10^{-3}$, along with $N = 400$ and $R_\mathrm{max} = 4\,R_\mathrm{out}$, where $R_\mathrm{out}$ is defined by the $2\sigma$ contour in the robust 0.5 image (similarly to \citealt{Andrews2021}). The parameter $\alpha$ determines the signal-to-noise ratio threshold at which the model stops fitting the data, while $w_\mathrm{smooth}$ helps suppress noisy oscillations. The hyperparameter $N$ sets the radial gridding resolution, and $R_\mathrm{max}$ defines the point beyond which \frank assumes zero emission.

It is not possible to obtain an accurate uncertainty on the intensity profile reconstructed by \frank, due to the inherently ill-posed nature of recovering a brightness profile from incomplete Fourier data \citep{Jennings2020, Jennings2022}. To estimate the uncertainties, we followed a bootstrapping approach similar to \cite{Carvalho2024}. We ran the \frank fit 500 times, randomly varying the geometrical parameters ($i$, PA, $\Delta$RA, and $\Delta$Dec) using Gaussian distributions centered on the best-fit values from \galario. We used a standard deviation of 1~deg for $i$ and PA, and 1/3 of the synthesized beam’s major axis for the positional offsets, corresponding to a centering accuracy of ${\sim}7$ mas. The resulting distribution of intensities at each radius was used to derive intensity uncertainties as the 16th and 84th percentiles. We then relied on these intensity uncertainties to assign errors to the dust disk extent estimated from visibility fitting (Table~\ref{Tab:rad_size}) and to the location and width of the gap used to infer the mass of the planet able to explain the substructures (Sect.~\ref{sect:planet_mass}).

Plots displaying the visibilities versus the deprojected baseline and the intensity radial profile, comparing data, galario model, and \frank model with their associated uncertainties, are shown in Figure~\ref{fig:uv_lambda}. Figure~\ref{fig:continuum_gallery} also presents the CLEAN image, the $2\pi$-swept \frank model, the CLEAN \frank model (the frank fit sampled at the same \uv locations as the observations and imaged with CLEAN), and the residuals between the data and the \frank model for robust value of 0.

The CLEAN image and the azimuthally averaged radial profile of the CLEAN data, as well as the azimuthally averaged radial profile of the CLEAN model, show no evidence for the presence of substructures. However, our analysis reveals that both the \galario\ and \frank\ models recover a faint ring at a radius of approximately $0.11\arcsec$ (15.4~au). A smooth intensity profile would not be consistent with the uncertainties shown for the \frank and \galario profiles. The presence of a ring is also supported by the real part of the visibilities, shown in Figure~\ref{fig:uv_lambda}, which deviates from a perfect Gaussian form and exhibits distinct structural features. In particular, the visibility profile crosses zero at approximately 2.75~M$\lambda$, becoming negative before returning to positive values at around 3.25~M$\lambda$. This behavior suggests the presence of substructures within the data, although a symmetric disk with a sharp outer edge could also produce negative visibilities.

To evaluate the statistical significance of the dust ring retrieved by \frank and \galario, in Appendix~\ref{sec:appendix_model_comparison} we compare these two models with a simpler \galario\ model including only a central Gaussian and no ring. We find that the \galario model with a ring is strongly favored by the Akaike Information Criterion (AIC; \citealt{Akaike1974}; $\Delta \text{AIC} = -24.4$), while the simpler model without the ring is preferred by the Bayesian Information Criterion (BIC; \citealt{Schwarz1978};  $\Delta \text{BIC} = +11.8$). This outcome is expected, as BIC penalizes model complexity more strongly, particularly given the large number of visibilities ($N \approx 1.3\times 10^6$).  Nonetheless, the high likelihood of the non-parametric \frank fit, which independently recovers a similar ring, points to the substructure being a real feature of the disk, rather than a product of model choice.

It is worth noting that the reason why the observed substructures in the visibilities are interpreted as a ring in both the \galario and \frank models may lie in the fact that both techniques are operating in 1D on azimuthally averaged visibilities. In reality, these substructures in the visibility profile may also imply a non-axisymmetric structure in the data. A deviation from axisymmetry is suggested by the presence of apparently coherent features in the imaginary part of the visibilities around 1.4 M$\lambda$ (lower-left panel in Fig.~\ref{fig:uv_lambda}). Moreover, the robust -0.5 image (Figure \ref{fig:continuum_gallery_r-05} in Appendix) reveals hints of non-axisymmetric structures, which become more apparent in the residuals obtained after subtracting the \frank model. In summary, while the current data quality suggests the presence of some form of substructure within the disk, higher resolution and higher sensitivity data will be necessary to confirm this finding.

\subsection{Size of disk: gas vs dust}\label{sec:gas-dust-size}
To determine the size of the dust and gas disks in our recent observations, we analyzed the radial intensity profile and calculated the cumulative flux from the disk center out to a given radius, $R_{\text{max}}$. We defined the radii $R_{68}$, $R_{90}$, and $R_{95}$ as the distances within which 68\%, 90\%, and 95\% of the total flux is enclosed, respectively. Consequently, we measured the disk radii in the image plane, obtaining $R_{68}$, $R_{90}$, and $R_{95}$ values of 0$\farcs$12 ($\sim$16.8 au), 0$\farcs$14 ($\sim$19.6 au), and 0$\farcs$15 ($\sim$21 au) for dust emission (see Table \ref{Tab:rad_size}). For the $^{12}$CO gas emission line, we derived radii of 0$\farcs$24 ($\sim$33.6 au), 0$\farcs$36 ($\sim$ 50.4 au), and 0$\farcs$40 ($\sim$56 au) from the moment 0 map. Flux uncertainties were estimated by multiplying the rms (Jy beam$^{-1}$), measured from an emission-free region of the image, by the square root of the number of beams encompassed by the aperture used for total flux extraction. Radii uncertainties were derived from the corresponding flux uncertainties where certain $\Delta\,R_{68}$ represents the aperture radius variation that encloses a flux change of  $\Delta\,F_{68}$, and similarly for $\Delta\,R_{90}$ and $\Delta\,R_{95}$. This procedure is adopted in several works (e.g., \citealt{Ansdell16-1,Ansdell18-1,Andrews2018}).

To further refine the disk size measurements, we performed visibility fitting for both the dust continuum and the $^{12}$CO gas emission. This method provides a model-independent approach to estimate disk radii, avoiding potential biases introduced by image reconstruction. The results (see Table \ref{Tab:rad_size}) confirm that gas disks are systematically larger than their dust counterparts, with gas-to-dust size ratios ranging from 6 to 8 in the visibilities, compared to 2 to 2.7 in the image plane. This difference arises because visibility fitting allows for higher resolution in the dust continuum, reaching approximately one-third of the synthesized beam size \citep{Sierra24b}, whereas in the image plane, extended gas emission is more difficult to recover due to noise, which can blend faint outer structures with the background. This effect leads to an underestimation of the gas disk size when measured directly from the image plane.

\section{Discussion}
\label{sec:p10_discussion}

\subsection{Substellar Mass of 2M0444}
We derived the dynamical mass of 2M0444 using two different methodologies: fitting in the image plane with \EDDY (Section \ref{sec:p10_SLAM}) with a value of $0.043^{+0.003}_{-0.002}\,M_{\odot}$ and fitting in the visibility plane (Section \ref{sect:fit_12co}) with a mass of $0.092_{-0.007}^{+0.019}$ M$_{\odot}$. The visibility-plane fit yields the most robust result, as it directly models the velocity field in each spectral channel and \EDDY does not take the beam convolution into account, which may result in smaller stellar mass especially given the small disk size compared to the angular resolution \citep{Aso15}. However, the visibility plane measurement is significantly higher than both the value obtained in the image plane and previous estimates from evolutionary models. In any case, the derived dynamical values, and especially their uncertainties, in this paper should be treated with caution given the low s/n of the data. 

A key discrepancy arises when comparing our visibility-plane mass with the expected effective temperature from evolutionary tracks. According to the latest models from \citet{Baraffe15-1}, an object of $\sim$0.09 M$_{\odot}$ should have an effective temperature of $\sim$3000 K, while the observed effective temperature of 2M0444 is 2838 $\pm$ 50 K \citep{Luhman04-1}, which is more consistent with the lower mass inferred from the image-plane fit. This suggests that additional factors, such as uncertainties in the evolutionary models or deviations from strict Keplerian motion in the disk, may be influencing our results.

In addition to the dynamical mass measurements, we revisited the disk mass of 2M0444. \citet{Riccietal14-1} reported a disk mass of $1.3 \pm 0.2$ M${\text{Jup}}$, assuming a dust-to-gas ratio of 100 and an absorption coefficient of 0.02 cm$^2$/g. Using the same parameters and assumptions (dust opacity of $\kappa\nu = 2$ cm$^2$/g, temperature of $\sim$12 K, and a dust-to-gas ratio of 100), we derived a similar but slightly higher disk mass of $1.36 \pm 0.2$ M$_{\text{Jup}}$, within the flux uncertainties. However, we note that the actual uncertainty in the disk mass is likely larger than reported, as dust opacities remain poorly constrained by more than a factor of 2. Considering the mass of the central object and assuming that all the disk material is eventually accreted onto the brown dwarf, 2M0444 would remain in the substellar regime based on the mass derived with \EDDY.

While this study does not focus on methodological comparisons, our analysis indicates that the dynamical mass of 2M0444 might be higher than previously thought. Nevertheless, we still classify 2M0444 as a brown dwarf, though we caution that its true mass may be near the substellar boundary which typically is assumed to be $0.075$ M$_{\odot}$ \citep{Chabrier2023}. Importantly, 2M0444 remains the lowest-mass object with a dynamical mass measurement from ALMA to date. Future high-sensitivity observations will help refine this result and clarify its classification.

\subsection{Evidence of efficient Radial drift}
Previous measurements of the dust disk around 2M0444 suggested an extended disk. Using visibility plane data, \citet{Testi2016} applied two different models for the dust distribution based on a two-layer disk structure. The first model, assuming a power-law distribution with a sharp outer boundary, estimated a disk size of $153^{+100}_{-40}$ AU, while the second model, incorporating an exponential taper, yielded a value of $49^{+48}_{-11}$ AU. The discrepancy between our measurements and those by \citet{Testi2016} is primarily due to differences in baseline length observation: their data reached a maximum baseline of 0.3 M$\lambda$, which limited the disk's resolution to $\sim$0$\farcs$5. In contrast, our observations extended up to 7.5 M$\lambda$ (see Figure \ref{fig:uv_lambda}), allowing for better resolution of the disk structure.

The radial extent of the CO gas measured in the visibility plane compared to the continuum emission differs by a factor of 2 at R$_{68}$ that can be explained from optical depth differences between the CO and the dust \citep{Trapman19}. However, when using the values derived from the visibility plane, which range between 6.3 and 7.7, the results align with the expected effects of radial drift \citep{Birnstiel14}. Over recent years, compact dust disks with extended CO disks have been more frequently observed around T-Tauri stars \citep{Facchini17,Trapman19,Facchini19}, especially as attention has shifted toward less luminous disks. The compact nature of these disks can be explained by the inward drift of pebbles toward regions of local pressure maxima, where the dust becomes decoupled from the gas.

\citet{Trapman19} found that an $R_{\text{gas}}/R_{\text{dust}} \geq 4$ ratio is a clear indication of efficient radial drift, rather than being related to optical depth effects. A notable example of radial drift in the stellar regime is CX Tau, which has a gas-to-dust size ratio of 5.4 when enclosing 68$\%$ of the flux and 3.9 for 90$\%$ of the flux \citep{Facchini19}. Similarly, \citet{Kurtovic21} studied the radial extent of gas and dust around six very low-mass stars ($<0.2,M_{\odot}$), all of which showed gas disks more than three times larger than the dust disks, indicating significant radial drift of millimeter-sized grains \citep{Pinilla2013, Zhu18}. This includes CIDA 7, which has a gas-to-dust disk ratio of 6, similar to 2M0444. These findings support the hypothesis that fast radial drift is associated with the inability to trap dust in the outer disk regions \citep{Long2019}, likely contributing to the compact nature of 2M0444.

\begin{table}
\footnotesize
\caption{ Geometrical parameters derived in Section \ref{sec:p10_cont_fit}.}
\label{Tab:geom_par}
\centering
\begin{tabular}{cc}
\hline \hline
Inclination & $49.9_{-0.5}^{+0.5}$ deg \\ 
Position Angle & $124.6_{-0.7}^{+0.7}$ deg\\
dRA & $0.0056_{-0.0001}^{+0.0001}$ arcsec\\ 
dDec & $-0.0050_{-0.0001}^{+0.0001}$ arcsec  \\
\hline
\end{tabular}
\end{table}

\begin{table}
\footnotesize
\caption{Radial size of the dust and $^{12}$CO emission line.}
\label{Tab:rad_size}
\centering
\begin{tabular}{ccccc}
\hline \hline
 &  & $R_{68}$($\arcsec$) & $R_{90}$($\arcsec$) & $R_{95}$($\arcsec$)\\
Image & $^{12}$CO & 0.242 $\pm$ 0.001 & 0.359 $\pm$ 0.004 & 0.402 $\pm$ 0.004\\ 
plane & Dust & 0.121 $\pm$ 0.001 & 0.144 $\pm$ 0.002 & 0.149 $\pm$ 0.002\\
 & Ratio & 2.00 & 2.49 & 2.69 \\
\hline
Visibility &$^{12}$CO & 0.360$^{+0.017}_{-0.014}$ & 0.506$^{+0.027}_{-0.029}$ & 0.545$\pm0.033$\\
plane &  Dust & 0.047$^{+0.002}_{-0.001}$ & 0.065$^{+0.006}_{-0.001}$ & 0.087$^{+0.011}_{-0.003}$ \\
 & Ratio & 7.7 & 7.8 & 6.3 \\
 \hline
\end{tabular}
Notes: Dust values from the visibility profile were measured from the \textit{Frank} fit. Dust values from the image plane were measured using a value of robust of 0.  $^{12}$CO values from the image plane were measured using the keplerian mask.
\end{table} 

\subsection{Maximum possible planet mass carving the gap} \label{sect:planet_mass}
One of the most widely accepted explanations for the presence of gaps in protoplanetary disks is the gravitational influence of an embedded planet \citep{Kanagawa17,Huang2018,Long2018}. \citet{Lodato2019} established empirical relationships between gap properties and the maximum potential planet mass responsible for carving the gap, based on two main assumptions: (1) the disks have low viscosity (with $\alpha < 0.01$), and (2) there is a one-to-one correspondence between a single gap and a single planet, meaning that no multiple planets or gaps created by a single planet are considered.

The planet's mass, $M_{p}$, can be estimated using the Hill radius ($R_{H}$) formula: \begin{equation}
    M_{p}= \frac{3M_{*}R_{H}^{3}}{R^{3}}
\end{equation}
where $M_{*}$ is the BD mass 0.043$^{+0.003}_{-0.002}$ M$_{\odot}$, derived in Section \ref{sec:p10_anal}, and R represents the gap's radial location, assumed to correspond to the planet's position. The Hill radius, defines the region where the planet's gravitational influence dominates over the other forces. It can be related to the gap's width ($\Delta$), which is the distance from the brightness minimum within the gap to the ring's peak, via the expression: \begin{equation}
    R_{H}= \frac{\Delta}{k}
\end{equation}
where k is a constant ranging from 4 to 8 depending on the disk parameters \citep{Dodson2011, Pinilla2012, Fung2016, Rosotti2016, Facchini2018-2, Lodato2019}. 

In the case of 2M0444, the gap width ($\Delta$) is 16.2$^{+7.8}_{-0.6}$ mas, and the radial distance to the center of the gap is 98.1$^{+4.2}_{-8.4}$ mas, as determined in Section  \ref{sec:p10_cont_fit}. If there is a gap-ring pair located in that position, then the planet able to carve that gap would have an estimated mass between 0.3 and 7.7 M$_{\oplus}$, assuming k values of 4 and 8, respectively.

CIDA 1 is the lowest mass star that exhibits substructures detected with ALMA. This very low-mass star, with a mass of 0.2 M$_{\odot}$ \citep{Curone22}, has a gap located at a distance of 9-10 au from the star. According to \citet{Curone22}, using empirical relations, this gap is likely carved by a giant gas planet with an estimated mass in the range from 1.4 to 4-8 M$_{\mathrm{Jup}}$. The formation of such a massive planet around CIDA 1 is thought to result from rapid disk fragmentation due to gravitational instability, which is considered incompatible with the process of pebble accretion.

In contrast, the estimated planet mass for 2M0444 is consistent with the core accretion scenario \citep{Pollack96-1}, where submicron-sized dust particles grow through collisions, eventually reaching kilometer-scale sizes. When dust particles grow to millimeter sizes, they tend to drift rapidly inward and may accrete onto the central object, preventing the formation of larger bodies like planetesimals. However, simulations by \citet{Payne07-1} demonstrated that Earth-mass planets can still form around brown dwarfs by rapidly developing a rocky core. This scenario may provide the most straightforward explanation for the substructures observed in 2M0444’s disk: an Earth-like planet could be carving the gap, formed swiftly via pebble accretion.

\section{Summary}
\label{sec:p10_summary}
We presented high-resolution ALMA observations at 0.89 mm of the Class II brown dwarf 2MASS J04442713+2512164, at a spatial resolution of $\sim$6.4 au. Our observations targeted the continuum emission as well as two CO isotopologues, $^{12}$CO (3-2) and $^{13}$CO (3-2). We summarize our findings below:

-The $^{12}$CO (3-2) emission line reveals a disk with Keplerian rotation. We derived the dynamical mass of the central object between 0.043 and 0.092 $M_{\odot}$. \\
-The best fit of the continuum visibilities suggests a hint of substructure, characterized by a pair of a gap and a ring located at 98.1$^{+4.2}_{-8.4}$ mas ($\sim$13.7 au) and 116.0$^{+4.2}_{-4.8}$\,mas ($\sim$16.2 au), respectively, with a gap width of 20 mas ($\sim$2.8 au). This hint of substructure is also evident in the visibility profile. \\
-If a planet is carving the possible gap, its mass is estimated to range from $0.3 M_{\oplus}$ to $7.7 M_{\oplus}$. \\
-The radial extent of the dust and gas has a ratio $>$6 indicating efficient radial dust drift toward the dust trap peaking at 15.4 au and this trap is sufficient to retain enough dust particles to have millimetre fluxes that are detectable. Without the trap, the dust would have been lost towards the star in less than 1 Myr. \\

While not necessarily exceptional, given the novelty of this high-resolution study, these findings open a new window into the processes governing dust dynamics in brown dwarf and very low-mass star disks. The potential link between the observed pressure bump and an unseen planetary gas companion warrants further investigation with even higher-resolution observations at Band 7 and the ALMA Wideband Sensitivity Upgrade to better constrain substructures and explore the diversity of very low-mass object disk architectures.

\begin{acknowledgments}
This paper makes use of the following ALMA data: ADS/JAO.ALMA$\#$2023.1.00158.S ALMA is a partnership of ESO (representing its member states), NSF (USA) and NINS (Japan), together with NRC (Canada), MOST and ASIAA (Taiwan), and KASI (Republic of Korea), in cooperation with the Republic of Chile. The Joint ALMA Observatory is operated by ESO, AUI/NRAO and NAOJ.
 A.S.M. acknowledges support from ANID / Fondo 2022 ALMA / 31220025.
 P.C. acknowledges support by the ANID BASAL project FB210003.
 A.R. has been supported by the UK Science and Technology Facilities Council (STFC) via the consolidated grant ST/W000997/1 and by the European Union’s Horizon 2020 research and innovation programme under the Marie Sklodowska-Curie grant agreement No. 823823 (RISE DUSTBUSTERS project). NH has been funded by the Spanish grants MCIN/AEI/10.13039/501100011033 PID2019-107061GB-C61 and PID2023-150468NB-I00. IdG thanks grants PID2020-114461GB-I00 and PID2023-146295NB-I00, funded by MCIN/AEI/10.13039/501100011033. MRS acknowledges financial support from FONDECYT (grant number 1221059). C.A.G. acknowledges support from the Joint Committee ESO-Government of Chile 2023.

\end{acknowledgments}

\vspace{5mm}
\facilities{ALMA}

\software{astropy \citep{astropy1,astropy2},  EDDY \citep{eddy}, \citep{bettermoments}, frankenstein \citep{Jennings2022}}

\appendix

\section{Additional tables}
Table \ref{Tab:p10_observations} shows the observing log, while Table \ref{tab:line_summary} shows the main properties of the $^{12}$CO and $^{13}$CO emission lines.
\begin{table*}
\footnotesize
\caption{Observing log}
\label{Tab:p10_observations}
\centering
\begin{tabular}{cccccccc}
\hline \hline
Dates & Config & Baselines & N$_\mathrm{ant}$ & Elev.& PWV & Calibrators & Check source \\
(1) & (2) & (3) & (4) & (5) & (6) & (7) & (8)\\
& & [m] & & [deg] & \\
\hline
12 Oct 2023 & C43-8 & 91-8282 &  44 & 40 & 0.6 &  J0510+1800, J0510+1800, J0433+2905 & J0440+2728 \\
13 Oct 2023 & C43-8 & 91-8547 &  44 &  33 & 0.8 & J0510+1800, J0510+1800, J0433+2905 & J0440+2728 \\
2 Nov 2023 &  C43-8 & 85-6582 &  43 &  40 & 0.7 & J0510+1800, J0510+1800, J0438+3004 & J0440+2728 \\
30 May 2024 &  C43-5 & 15-1261 & 45 & 38 & 0.7  & J0423-0120, J0423-0120, J0438+3004 & J0440+2728 \\
\end{tabular}
Notes:(1) Observing date. (2) Antenna configuration. (3) Baseline length. (4) Number of antennas. (5) Elevation. (6) Precipitable water vapour. (7) From left to right quasars used for calibrating the bandpass, amplitude and phase variations. (8) checking the phase transfer. 
\end{table*} 

\begin{table*}
\footnotesize
\caption{Main properties of the detected molecular lines.}
\label{tab:line_summary}
\centering
\begin{tabular}{cccccccc}
\hline
\hline
 Frequency & Molecule & Transition & robust  & Beam size & $\Delta$v$^{a}$ &  Integrated intensity$^{b}$ & Peak intensity \\

 [GHz]       &    & &    & [$\arcsec\times\arcsec$] & [km s$^{-1}$] & [mJy km s$^{-1}$] & [mJy beam$^{-1}$] \\
\hline 
330.588 & $^{13}$CO & 3--2 & 2 &  0.074$\times$0.056 &  4.43 &  136 $\pm$ 15 & 21.6 $\pm$ 5.4\\ 
345.796 & $^{12}$CO & 3--2 & 2 & 0.071$\times$0.054 & 7.20 & 427 $\pm$ 43 & 51.5 $\pm$ 7.2\\ 
\hline
\end{tabular} 
\begin{flushleft}
$^{a}$ Velocity width of the detected line measured with more than 3$\sigma$ detection.  \\
$^{b}$Integrated intensity over the whole emission area, obtained from a 3$\sigma$ contour over the moment 0 map. \\ 
Integrated intensity and peak intensity uncertainties consider a 10$\%$ absolute calibration uncertainty and the noise in the images.
\end{flushleft}
\end{table*}

\section{Additional figures}
\begin{figure*}
\includegraphics[width=0.90\textwidth]{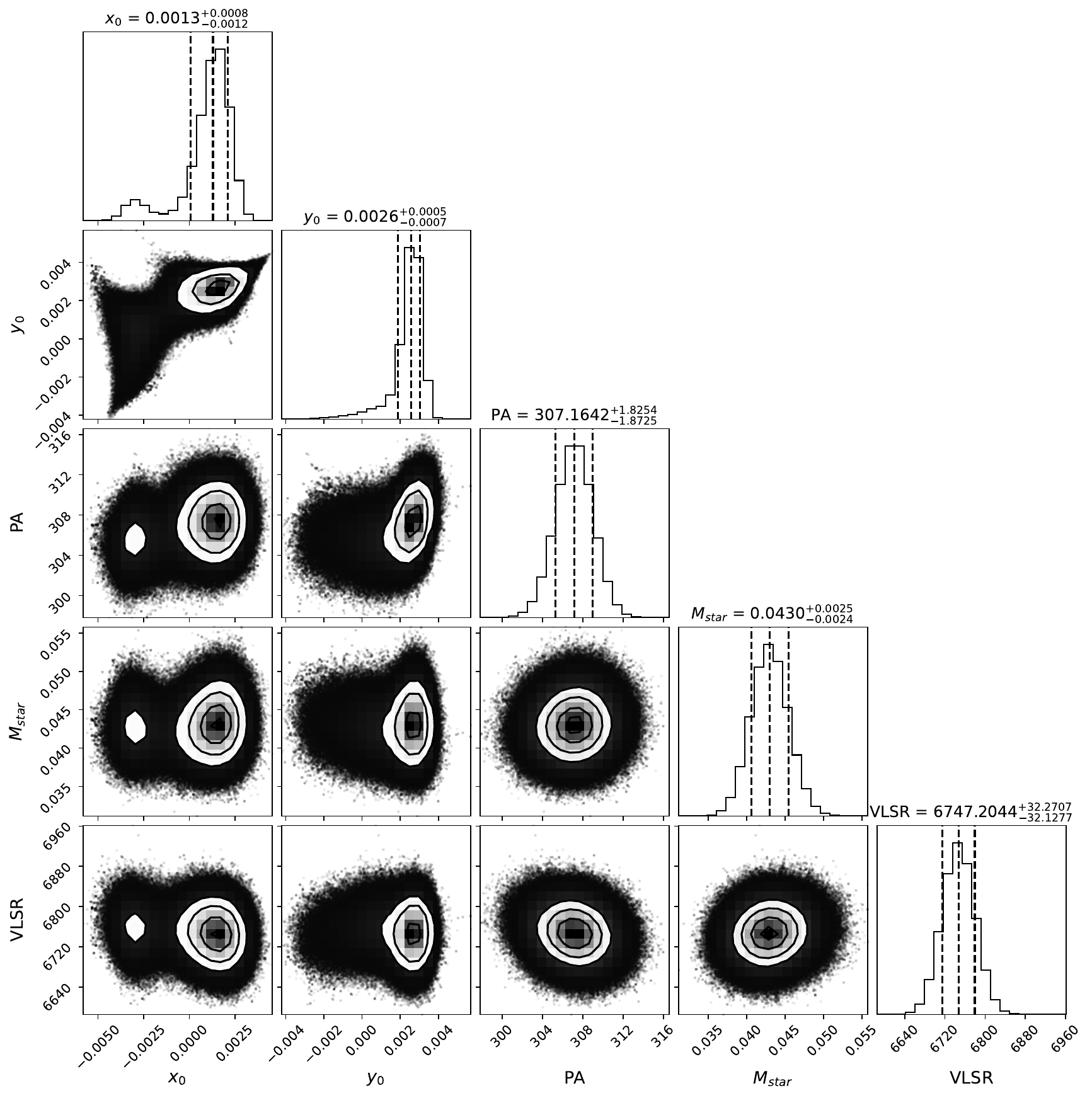}
\caption{Corner plot of the \EDDY fitting. x$_{0}$ and y$_{0}$ are the shifted coordinates in mas respect to the image center.  PA is the position angle in degrees. M$_{star}$ is the mass of the central object in solar masses. VLSR is the local standard rest velocity of the source in km s$^{-1}$. Dashed lines are the 16 and 84 percentiles.}.\label{cornerplot_eddy}
\end{figure*}

\begin{figure*}
\includegraphics[width=0.90\textwidth]{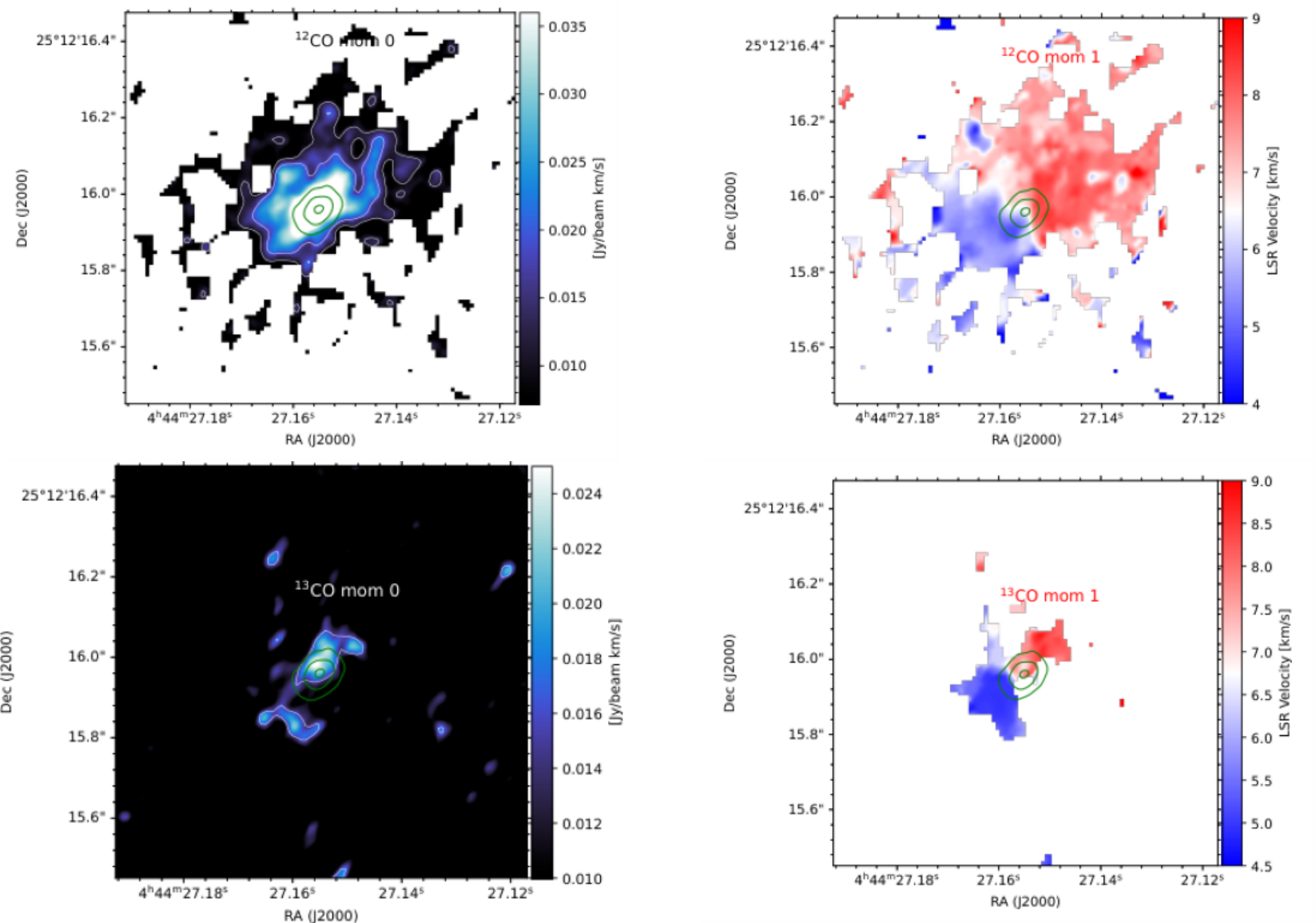}
\caption{Moment maps of the detected CO isotopologues. The top row shows $^{12}$CO (3-2), and the bottom row shows $^{13}$CO 3-2). The left column shows the integrated intensity map, and the right column shows the intensity-weighted velocity map.  Green contours represent the disk continuum emission with 10, 40 and 80 times the rms. White contours represent the 3, 5 times the rms (1$\sigma$ = 3.6 mJy beam$^{-1}$ km s$^{-1}$) for the $^{12}$CO (3-2) moment 0 map; 3 times the rms for $^{13}$CO (3-2) moment 0 (1$\sigma$ = 5.0 mJy beam$^{-1}$ km s$^{-1}$).} \label{momentos_no_keplerianos}
\end{figure*}

\begin{figure*}
\includegraphics[width=1.0\textwidth]{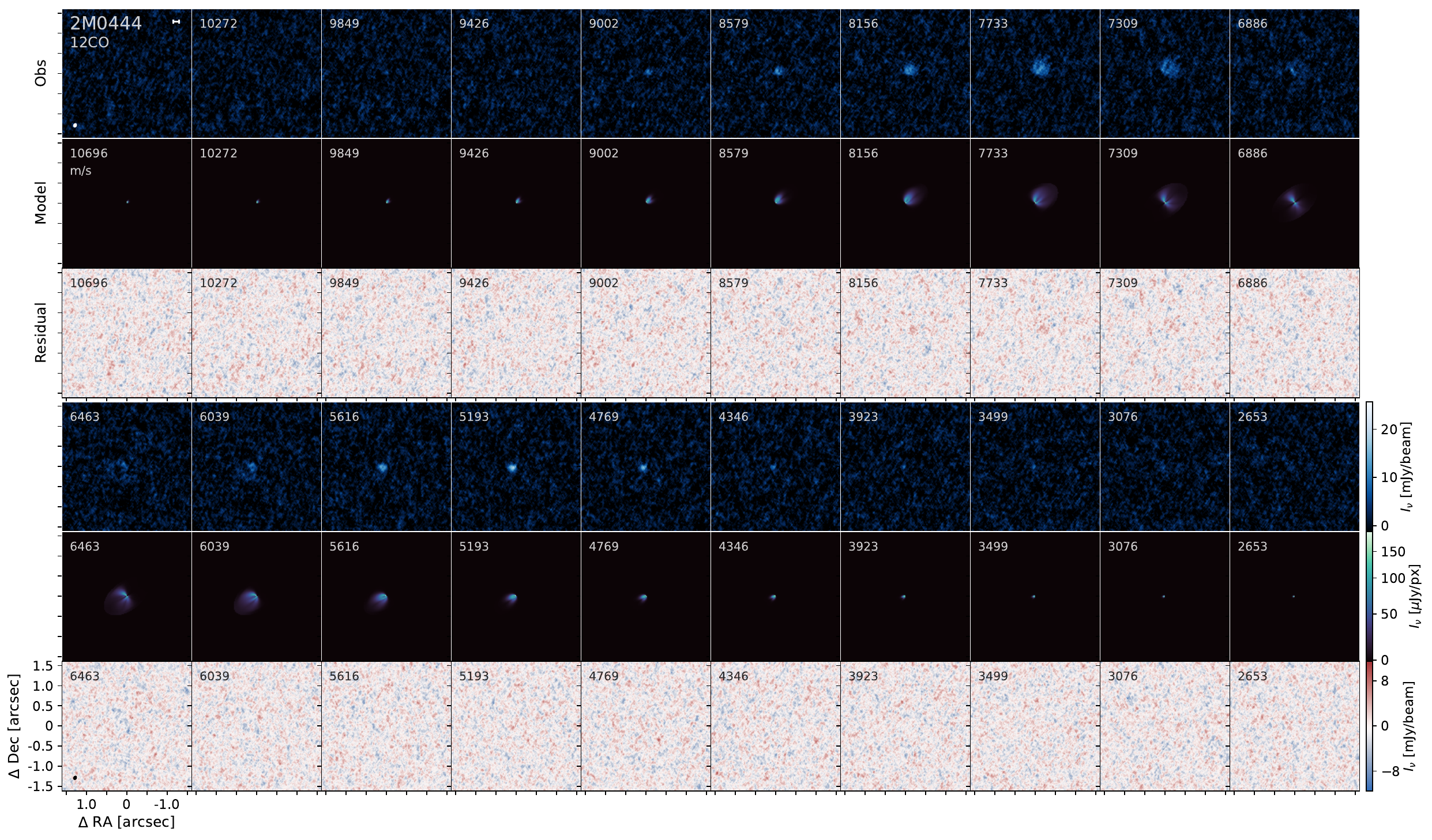}
\caption{Comparison of the observed $^{12}$CO emission (top rows), the highest likelihood brightness distribution model (middle rows), and the residuals after subtracting the visibilities of the model to the observation (bottom rows). The numbers in the top left of each panel show the LSRK velocity of each channel. }\label{app:fig:vismodel_12co}
\end{figure*}

\begin{figure*}
\includegraphics[width=\textwidth]{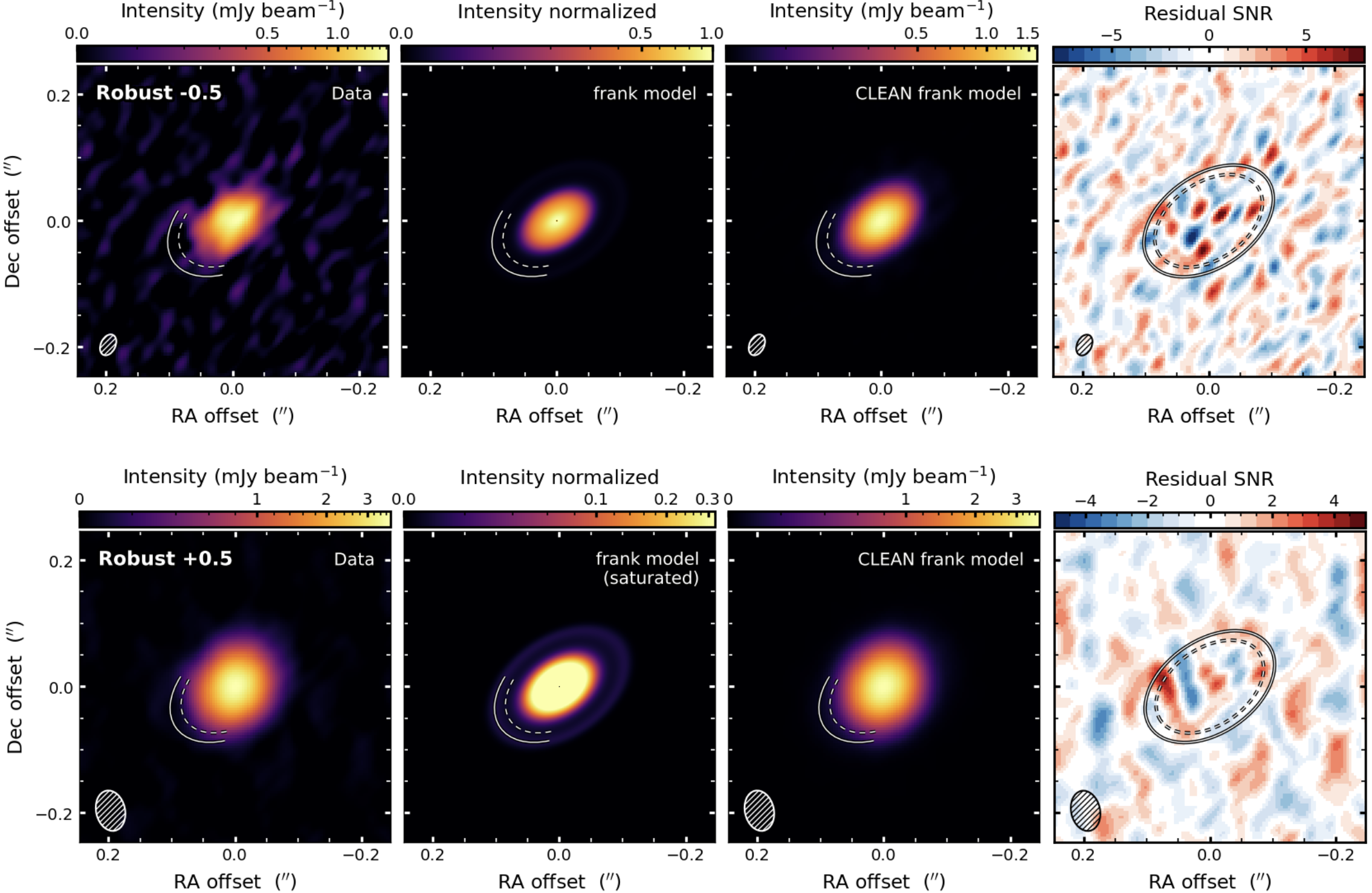}
\caption{CLEAN data, \frank model, \frank CLEAN image, and residual between data and \frank model imaged with robust -0.5 and 0.5. Dashed contours indicate the [3, 5, 10, 20, 50]$\sigma$ levels of the CLEAN data image.}
\label{fig:continuum_gallery_r-05}
\end{figure*}

Figure \ref{cornerplot_eddy} presents the corner plot of the EDDY fitting. Figure \ref{momentos_no_keplerianos} shows the moment maps for the $^{12}$CO and $^{13}$CO emission lines without Keplerian masking. Figure \ref{app:fig:vismodel_12co} shows the highest likelihood brightness model for the $^{12}$CO emission, fitted by comparing its visibilities to those observed by ALMA. Figure \ref{fig:continuum_gallery_r-05} is similar to Figure \ref{fig:continuum_gallery} using robust 0.5 and -0.5.

\begin{figure*}
\centering
\includegraphics[width=\textwidth]{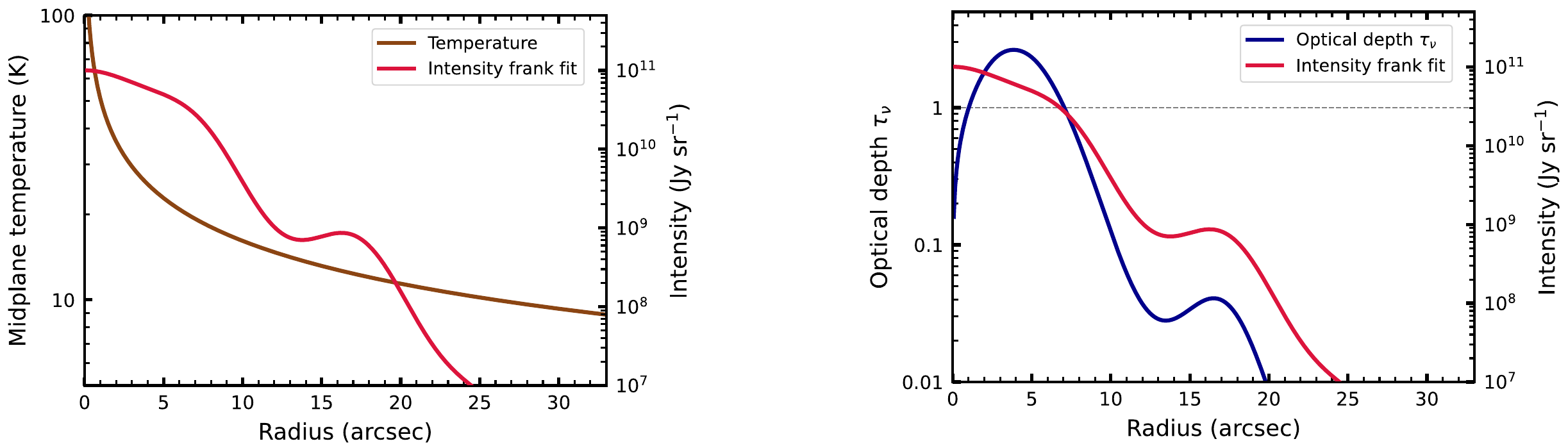}
\caption{Dust midplane temperature profile (left) and optical depth profile (right) computed as in \cite{Huang2018} compared to the intensity radial profile of the best-fit \frank model.}
\label{fig:temp_tau}
\end{figure*} 

\begin{figure*}
\centering
\includegraphics[width=0.90\textwidth]{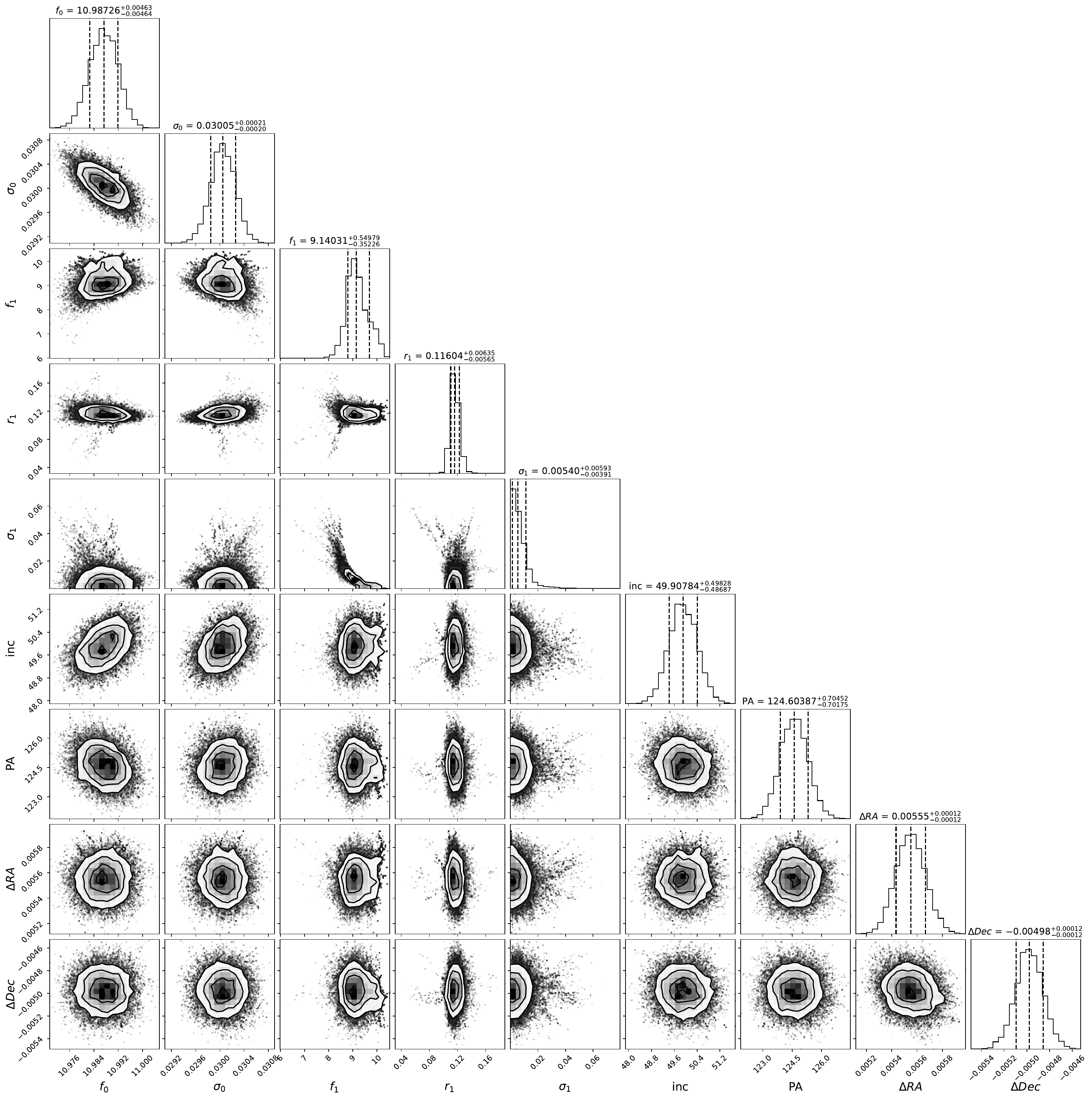}
\caption{Corner plot of the \galario MCMC run assuming a model with a central Gaussian plus a Gaussian ring}, showing the position of the walkers in the last 1000 steps.
\label{fig:galario_cornerplot}
\end{figure*} 

\begin{figure}
\centering
\includegraphics[width=0.5\columnwidth]{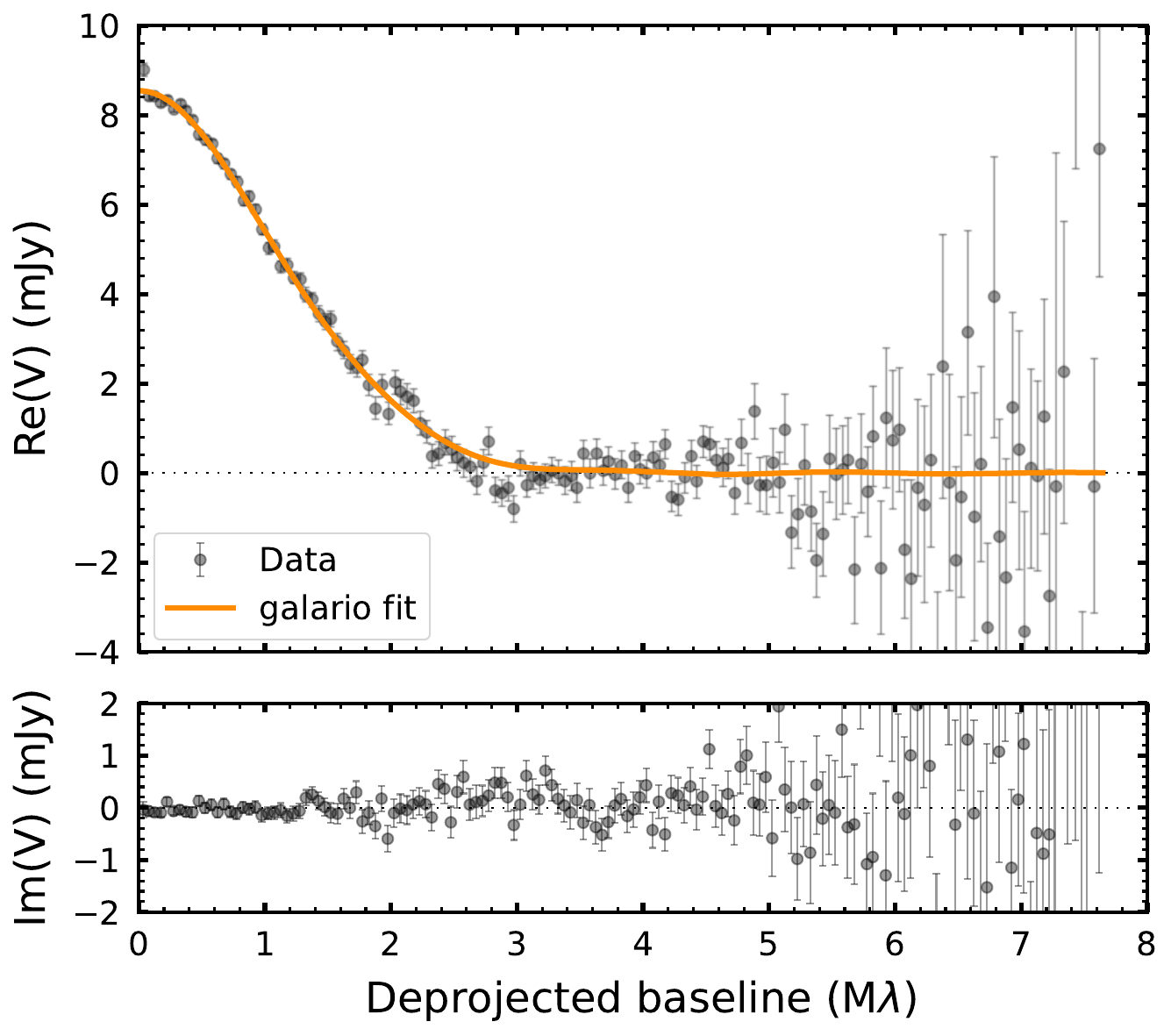}
\caption{Real and imaginary parts of the recentered and deprojected visibilities, azimuthally averaged into 50 k$\lambda$-wide bins, shown as a function of the deprojected baseline length for the data (gray points) and the best-fit model from \galario (orange line). Note that the \galario best-fit model is shown only for the real part, as the imaginary part is not fitted due to the assumption of an axisymmetric model.} \label{fig:galario_profile}
\end{figure} 

\begin{figure*}
\centering
\includegraphics[width=0.90\textwidth]{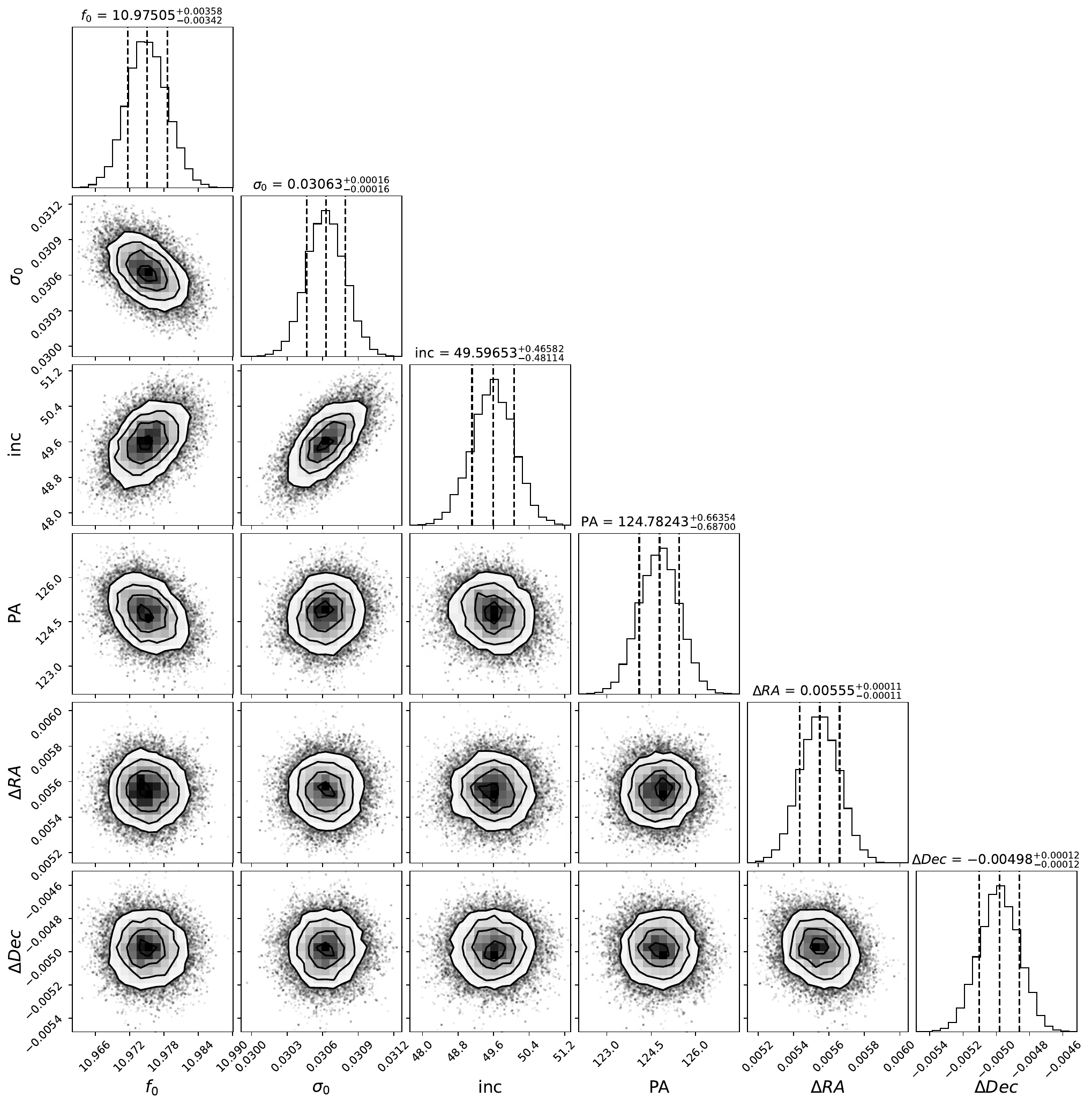}
\caption{Corner plot of the \galario MCMC run assuming a model with a central Gaussian only, showing the position of the walkers in the last 1000 steps.}
\label{fig:galario_cornerplot_simpleGauss}
\end{figure*} 

\section{Temperature and optical depth profile}  \label{sect:dust_mass_tau}
To obtain a rough estimate of the optical depth of the dust emission, we applied the procedure outlined by \cite{Huang2018}, which was also used in \cite{Facchini19}. First, we adopted a midplane radial temperature profile defined as $T_\mathrm{mid}(R) = [(\varphi L_\star) / (8\pi R^2\sigma_\mathrm{SB})]^{0.25}$, where $\sigma_\mathrm{SB}$ is the Stefan-Boltzmann constant, $L_\star$ is the stellar luminosity, and $\varphi$ represents the flaring angle, which we set to 0.02. Using this temperature profile, we computed the radial profile of the optical depth ($\tau_\nu$) based on the relation $I_\nu(R) = B_\nu(T_\mathrm{mid}(R)) [1 - \exp(-\tau_\nu(R))]$, where $I_\nu(R)$ is taken from the best-fit \frank model. The resulting profiles of the temperature and optical depth are presented in Figure~\ref{fig:temp_tau} . The optical depth profile indicates that the disk is optically thick within a radius of ${\sim}$0.05~arcsec (${\sim}$7~au).
\label{appendix:temp_tau}

\section{\galario fit}
Here we present the results of the \galario fit run using 100 walkers and 10000 steps, assuming an intensity model composed of a central Gaussian and an additional Gaussian ring (see Sect. \ref{sec:p10_cont_fit}). Figure~\ref{fig:galario_cornerplot} presents the resulting corner plot while Figure~\ref{fig:galario_profile} shows the deprojected visibilities and the best-fit model.

\section{Statistical comparison of dust substructure models}
\label{sec:appendix_model_comparison}

To assess the statistical significance of the continuum substructure, we compare the non-parametric fit obtained with the \frank code and the parametric fit using \galario with a central Gaussian plus a ring (presented in Sect. \ref{sec:p10_cont_fit}) with a simpler \galario fit including only a central Gaussian. The parametric form of the latter model is $I(R) = f_0 \exp(-R^2 / 2\sigma_0^2)$ and its posterior distributions are shown in the corner plot in Fig.~\ref{fig:galario_cornerplot_simpleGauss}.

We compare the \frank fit to the central Gaussian model by evaluating the maximum log-likelihood achieved by each fit. The \frank model yields the highest log-likelihood, with a difference of $\Delta \ln \mathcal{L}_{\text{max}}=20.7$ relative to the central Gaussian model, indicating a substantially better fit to the data. Since the two \galario models are parametric, we further compare them using statistical model selection criteria. We first compute the Akaike Information Criterion (AIC; \citealt{Akaike1974}), defined as $\text{AIC} = 2k - 2\ln(\mathcal{L}_{\text{max}})$, where $k$ is the number of free parameters. The Gaussian+Ring model is strongly favored by the AIC ($\Delta\text{AIC}=-24.4$), suggesting that the inclusion of the ring component improves the fit enough to justify the additional complexity. We also compute the Bayesian Information Criterion (BIC; \citealt{Schwarz1978}), defined as $\text{BIC} = k \ln N - 2 \ln(\mathcal{L}_{\text{max}})$, where $k$ is the number of free parameters and $N$ is the number of data points (i.e., the number of visibilities in our case, $\mathbf{N\approx1.3\times 10^6}$). The BIC favors the simpler Gaussian model ($\Delta\text{BIC}=+11.8$), as expected given its stronger penalization of model complexity.

These results imply that while the Gaussian+Ring model is supported under AIC, the evidence for the additional substructure is less compelling under BIC. The high likelihood of the frank model, which independently recovers similar substructure, supports the interpretation that the ring-like feature is real and not an artifact of a particular model family.

\bibliography{sample631}{}
\bibliographystyle{aasjournal}
\end{document}